\documentclass[aps,floatfix,nofootinbib,preprint]{revtex4}

\usepackage{graphicx}
\usepackage{epic}
\usepackage{eepic}
\usepackage{latexsym}
\usepackage{amssymb,amsmath}

\newcommand{\eq}[1]{(\ref{#1})}
\newcommand{\be}{\begin{equation}}
\newcommand{\ee}{\end{equation}}
\newcommand{\bea}{\begin{eqnarray}}
\newcommand{\eea}{\end{eqnarray}}

\newcommand{\vs}[1]{\vspace{#1 mm}}
\newcommand{\hs}[1]{\hspace{#1 mm}}

\def\a{\alpha}
\def\b{\beta}
\def\cc{\gamma}

\def\d{\delta}

\def\e{\epsilon}

\def\fr{\frac}

\def\m{\mu}
\def\n{\nu}

\def\r{\rho}
\def\s{\sigma}

\def\del{\partial}

\let\bm=\bibitem
\def\nn{\nonumber}

\begin{document}

\title{Time Evolution in Canonical Quantum Gravity is Trivial} 

\author{Ali Kaya}

\email[]{alikaya@tamu.edu}
\affiliation{\vs{3}Department of Physics and Astronomy, Texas A\&M University, College Station, TX 77843, USA \vs{10}}

\begin{abstract}

\vs{5}

The Wheeler-DeWitt (WdW) equation does not describe any explicit time evolution of the wave function, and somehow related to this issue, there is no natural way of defining an invariant inner product that provides a viable probability interpretation. We show that both of these difficulties are solved in a covariant canonical formulation of general relativity where the configuration space is extended by introducing the embedding coordinates as dynamical variables. The formalism describes the evolution of the wave function from one spacelike slice to another, but as in the case of spatial diffeomorphisms this is simply implemented by a coordinate change in the wave function. We demonstrate how  the time evolution equation disappears after gauge fixing that removes the embedding coordinates. These findings indicate that the time evolution is trivial in a background independent formulation of quantum gravity. 

\end{abstract}

\maketitle

The canonical analysis applied to the standard Einstein-Hilbert action shows that the Hamiltonian of general relativity becomes a linear combination of the first class constraints and hence weakly vanishes (in this work we will assume no boundary terms arise because either the space is closed or suitable boundary conditions are imposed ensuring their absence). Here, the spacetime covariance is broken since one splits the coordinates $X^\m=(t,x^i)$ and uses the Arnowitt-Deser-Misner (ADM) decomposition $g_{\m\n}\to (N, N^i, h_{ij})$. In the quantum theory, the wave function obeys
\bea
\Phi_i(x)\Psi=0,\nn\\
\Phi(x)\Psi=0,\label{wdw} 
\eea
where the so called momentum and Hamiltonian constraints, $\Phi_i(x)$ and $\Phi(x)$, depend locally only on $h_{ij}$ and its conjugate momentum $P^{ij}$ (the lapse $N$ and the shift $N^i$ become Lagrange multipliers). Eq. \eq{wdw} yields a set of functional differential equations for the wave function $\Psi[h_{ij}]$ and there seems to be no Schr\"{o}dinger equation involving $\del\Psi/\del t$. It looks like the canonical quantization does not describe the time evolution of the states and hence gives a timeless theory, which makes the  interpretation of the wave function puzzling. 

There are various proposals to solve the  problem of time in quantum gravity, however  a satisfactory solution  has yet to be found (see \cite{k1}  for an extensive review).  For example, a standard approach is to  deparametrize the theory by gauge fixing, after which a field (or a combination of the fields) in the theory becomes an emergent time parameter that keeps track of time evolution  (the remaining degrees of freedom are quantized in the standard way). Although such a procedure seems  straightforwardly applicable, e.g.,  in the path integral quantization \cite{bar1,bar2},  the gauge fixing generally breaks down and consequently the field ceases to be an honest time variable. Only in some special models one can find a suitable variable that can play the role of time. For example, the scale factor of the universe can be used as an emergent time in a flat minisuperspace quantum cosmology of a self-interacting scalar field having a positive definite potential; yet  it fails to be an honest time variable if the potential is not positive definite \cite{ak1, ak2}. Thus, deparametrization does not offer a satisfactory general solution to the problem of time and other proposals have similar issues. 

A somehow related matter, which is rarely discussed in the literature, is to define an inner product providing a viable probability distribution over the field variables. This is, in particular, contingent on determining the proper degrees of freedom and thus on identifying the time variable, which is not dynamical in quantum mechanics. Another difficulty that is often overlooked in the discussion is that the WdW equation is structurally similar to the Klein-Gordon equation; therefore, the same historical puzzle of defining a positive definite invariant inner product applies here. This either leads to third quantization where the wave function $\Psi$ turns to an operator or to unorthodox solutions like the one discussed in \cite{must1, must2}, which uses pseudo-Hermitian operators. Either possibility has considerable interpretation issues. 

It is well known that the momentum constraint $\Phi_i(x)\Psi=0$ ensures the spatial diffeomorphism invariance of the wave function $\Psi[h_{ij}]=\Psi[h_{ij}+\d h_{ij}]$, where $\d h_{ij}={\cal L}_{k^l} h_{ij}=k^l \del_l h_{ij} + \del_i k^l h_{lj}+\del_j k^l h_{li}$, ${\cal L}$ denotes the Lie derivative and $k^i$ is a spatial vector (here one should assume the ordering where  the momentum operators are placed in the rightmost position in $\Phi_i(x)\Psi=0$). In a full  covariant  description (e.g. in the Lagrangian formulation), a time translation also corresponds to a coordinate change but obviously the usual canonical analysis breaks the general covariance. One would then hope that a covariant canonical analysis might improve our understanding of time in  quantum gravity. 

It is possible to define a covariant phase space by using the fact that each point in the phase space actually corresponds to a classical solution. Given any Lagrangian, one can define a covariant symplectic structure in the space of solutions that coincides with the standard symplectic structure in the phase space when restricted to the initial value surface \cite{cw,lw}. Although this construction provides a covariant canonical description,  it is not clear how it  might be useful in quantization since it is difficult to define and interpret quantum states on the space of solutions. As a novel alternative approach, following \cite{ik1, ik2}, we have shown in \cite{aks1,aks2} that the canonical analysis of general relativity can be carried out in a spacetime covariant way by extending the configuration space with the so called embedding coordinates. Moreover, the extended theory can be quantized in a straightforward way and the formalism allows one to apply local time translations, a salient feature that plays an important role in resolving the problem of time as we will discuss below. 

To outline our construction, let $M$ be the spacetime manifold having (local) coordinates $X^\m$ and let $g_{\m\n}$ be a Lorentzian metric on $M$. We extend the configuration space by adding the Lorentzian foliations, which can be parametrized by the embedding maps $X^\m=X^\m(y^\a)$ where $y^\a=(\tau,y^i)$ and $\tau$ is an intrinsic time variable. The embedding coordinates must obey $g_{\m\n}X'^\m X'^\n<0$ and $\det \,( g_{\m\n}\del_\a X^\m \del_\b X^\n)<0$, where the prime denotes the $\tau$ derivative. We define the configuration space to be the collection of all fields $(X^\r,g_{\m\n})$ obeying the above conditions. Note that the embedding coordinates $X^\m(y^\a)$ can be viewed to describe the motion of a 3-brane in a 4-dimensional spacetime, where $X^\m$ and $y^\a$ are the so called target-space and world-volume coordinates, respectively. 

We now consider the following action 
\be\label{pga} 
S=\fr12 \int d^4y \,\sqrt{-\cc}\, R(\cc),
\ee
where $\cc_{\a\b}$ is the composite/induced metric given by 
\be\label{cc}
\cc_{\a\b}=\del_\a X^\m \del_\b X^\n g_{\m\n}, 
\ee
$\cc=\det \cc_{\a\b}$ and $R(\cc)$ is the Ricci scalar of $\cc_{\a\b}$.  The theory has two field variables $g_{\m\n}$ and $X^\m$, which are assumed to depend on  $y^\a$. Compared to general relativity, there is a larger gauge symmetry; the first is generated by a (world-volume) vector $k^\a$ so that 
\bea
&&\d X^\m=k^\a\del_\a X^\m,\nn\\
&&\d g_{\m\n}=k^\a \del_\a g_{\m\n}\label{wvd}
\eea
which induces $\d \cc_{\a\b}={\cal L}_k \cc_{\a\b}=k^\cc \del_\cc \cc_{\a\b}+\del_\a k^\cc  \cc_{\cc\b}+\del_\b k^\cc \cc_{\cc\a}$. The second is generated by a (target-space) vector $l^\m$ so that 
\bea
&&\d X^\m=l^\m,\nn\\
&&\d g_{\m\n}=-\del_\a l^\r X_\m^\a  g_{\r\n}  -\del_\a l^\r X_\n^\a  g_{\r\m},  \label{tsd}
\eea
which implies $\d \cc_{\a\b}=0$ (here $X_\m^\a$ is the matrix inverse of $X_\a^\m=\del_\a X^\m$). As discussed in \cite{aks1}, \eq{pga} is indeed equivalent to general relativity as can be seen by partially fixing the gauge $X^\m=\d^\m_\a y^\a$, which eliminates the embedding coordinates and leaves the usual diffeomorphism invariance as the residual gauge symmetry. 

The canonical analysis of the above theory can be carried out while preserving the spacetime covariance (on the other hand, the world-volume covariance is broken since one has to split $y^\a=(\tau,y^i)$). The canonical pairs can be identified as $(X^\m,P_\n)$ and $(g_{\m\n}, P^{\r\s})$, where the conjugate momenta can be obtained by varying the action \eq{pga} so that  $P_\m=\d S/\d X'^\m$ and $P^{\m\n}=\d S/\d g'_{\m\n}$ (as above the prime denotes the $\tau$ derivative).  One can show that the Hamiltonian vanishes identically (recall that we ignore the surface terms in this work), which should be contrasted to the usual canonical construction in which the Hamiltonian vanishes weakly \cite{aks1}. 

The details of the canonical analysis is a bit cumbersome but straightforward to carry out. There are primary constraints related to the non-invertibility of the momentum-velocity map and some secondary constraints arise from their Poisson brackets. At this point, it is useful to introduce $n^\m$ the future pointing unit normal vector of the foliation that obeys $n_\m \del_i X^\m=0$ and $n^\m n^\n g_{\m\n}=-1$, where $\del_i=\del/\del y^i$. One can also define the induced 3-dimensional metric 
\be
h_{ij}=\del_i X^\m \del_j X^\n g_{\m\n}, 
\ee
its determinant $h=\det(h_{ij})$, covariant derivative $D_i$ and Ricci scalar $R^{(3)}(h_{ij})$.  The momentum $P^{\m\n}$ can be projected out as 
\bea
&&P_{ij}=\del_i X^\m \del_j X^\n P_{\m\n},\nn\\
&&P=P^{ij}h_{ij}.
\eea	
Using these definitions, the full set of the constraints can be written as \cite{aks1} 
\bea
&&n_\n P^{\m\n}=0,\nn\\
&&\del_i X^\m P_\m+(\del_i g_{\m\n}) P^{\m\n}=0,\nn\\
&&n^\m P_\m + (\d_n g_{\m\n})\, P^{\m\n}=0,\label{rgrc}\\
&&\Phi_i= D_j \left(\fr{1}{\sqrt{h}} P^j{}_i\right)=0,\nn\\
&&\Phi= \fr{2}{\sqrt{h}}\left[P^{ij}P_{ij}-\fr12 P^2\right]-\fr12 \sqrt{h}R^{(3)}=0,\nn
\eea
where $(i,j)$ indices are manipulated by the  metric $h_{ij}$ and $\d_n g_{\m\n}$ denotes  the transformation given in \eq{tsd} with $l^\m=n^\m$. Not surprisingly, there are more constraints here since the gauge symmetry is enlarged. 

The second and the third constraints in \eq{rgrc} generate the symmetry \eq{wvd}.  It is possible to rewrite them (after some algebra)  in the following form
\be\label{covcons}
P_\m+2 P^{ij}\,D_i \left(\del_jX^\n \, g_{\m\n}\right)=0,
\ee
which can be shown to generate \eq{tsd}. The former can be obtained from \eq{covcons} by contracting with $\del_i X^\m$ and $n^\m$. Although their generators are related by  a (field dependent) contraction, it is crucial to note that \eq{wvd}  and \eq{tsd}  are two different symmetries; the former is generated by a world-volume vector $k^\a$ as its parameter and the latter is generated by a spacetime vector $l^\m$. Note also that  in general a symmetry generator simply multiplied by {\it a field dependent term} will not be a symmetry generator anymore; this happens in our case since there are actually two different symmetries. 

It is not difficult to quantize the system canonically:  The configuration space variables  $X^\m=X^\m(y^i)$ and $g_{\m\n}=g_{\m\n}(y^i)$ are defined on a constant $\tau$-surface corresponding to a spacelike slice in the spacetime.  To satisfy the canonical commutation relations the momentum operators can be defined as functional derivatives $P_\m=-i\d/\d X^\m$ and $P^{\m\n}=-i\d/\d g_{\m\n}$  acting on a  wave-function $\Psi$, which is a functional $\Psi=\Psi[X^\m,g_{\m\n}]$. 

A standard technical issue about the equations in quantum mechanics that follow from \eq{rgrc} is the ordering ambiguity. As it turns out, the first four constraints in \eq{rgrc} have clear geometrical  implications when the momenta are placed in the rightmost position. Assuming this natural ordering, let us now discuss the consequences of \eq{rgrc}:

(i) The first constraint in \eq{rgrc} implies  that the wave-function takes the form $\Psi=\Psi[X^\m,h_{ij}]$, i.e. if one  decomposes the metric with respect to  the tangent and  non-tangent surface components (by projecting out with $\del_i X^\m$), $\Psi$ becomes independent of the latter. This is the covariant version of the statement that the wave function does not depend on the lapse and the shift. 

(ii) The second constraint in \eq{rgrc} ensures the invariance of $\Psi$ under a spatial diffeomorphism generated by $\d X^\m=k^i \del_i X^\m$, where the metric must be transformed according to \eq{wvd}. 

(iii) The third constraint in \eq{rgrc} determines the change in $\Psi$ when the initial spacelike slice is deformed along its normal direction $\d X^\m= n^\m$. The new state can be determined by changing the metric in the wave function according to the gauge transformation \eq{tsd}. 

(iv) As in the usual WdW quantization, the fourth constraint in \eq{rgrc} ensures the spatial covariance of the wave function, i.e. $\Psi[X^\m,h_{ij}]=\Psi[X^\m,h_{ij}+\d h_{ij}]$, where $\d h_{ij}={\cal L}_k h_{ij}$ for an arbitrary $k^i$. 

(v) As we will discuss below, the fifth constraint in \eq{rgrc} restricts the degrees of freedom in the theory. 

\begin{figure}
	\centerline{\includegraphics[width=9cm]{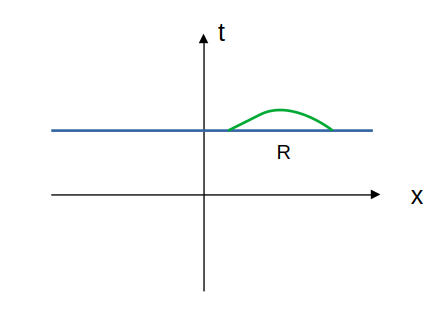}}
	\caption{An initial spacelike constant-$t$ slice and its deformation along the normal vector field in a local region R. The deformation should yield a new spacelike slice.} 
	\label{fig1}
\end{figure}

We thus see by the point (iii) above that in the extended theory there exists an equation dictating the time evolution of the states. Unlike the usual Schr\"{o}dinger equation, which describes the change of a state under a rigid time translation, the new equation is local, i.e. one can consider an arbitrary position dependent deformation of an initial slice along its normal direction $\d X^\m=\e(y^i)n^\m$, where $\e(y^i)$ is an arbitrary (infinitesimal) parameter, see Figure \ref{fig1}. Crucially, the effect of this transformation is purely determined by the covariance; the spatial diffeomorphisms and time translations gain the same status as gauge transformations. 

To dwell on this point more, we note that the primary constraint following from the  action \eq{pga}  is actually given by
\be
n^\m P_\m + (\d_n g_{\m\n})\, P^{\m\n}+\Phi=0. \label{ham-primary}
\ee 
Such an equation having the same structure also arises in a self interacting scalar field theory extended by the embedding coordinates, where $\Phi$ is the Hamiltonian density of the scalar field \cite{aks1}. Thus, at this point based on \eq{ham-primary}, the time evolution in quantum gravity and that of a quantum scalar field theory appear to be the same, where $\Phi$ generates {\it local time translations.} However, the commutator of \eq{ham-primary} with the first equation in \eq{rgrc}, which is another primary constraint, implies $\Phi=0$. As a result, the time evolution in canonical quantum gravity loses its dynamical character and becomes solely determined by the covariance. 

It is important to emphasize that the operator $n^\m P_\m$ generates a real deformation in the spacetime like in Fig. \ref{fig1} and does not simply correspond to a repartmentalization of the same initial surface since $P_\m$ generates the translation of the spacetime coordinate $X^\m$. A possible factor which may mislead one to think on the contrary is  $\d\cc_{\a\b}=0$  under \eq{tsd}, however this simply means the brane in our brane-anology is not dynamical. Indeed, as shown in \cite{aks1},  \eq{ham-primary} yields the standard Schr\"{o}dinger  equation for a self-interacting scalar field where  $n^\m P_\m$ becomes the operator $\del/\del t$. As pointed out above, the main difference of general relativity is the extra constraint $\Phi=0$, yet the kinematics are completely identical in the two theories. 

To complete the quantization, one should also introduce an invariant inner product that yields, for a given wave function, a viable probability distribution over the physical degrees of freedom. Such an inner product with desired properties has been defined in \cite{aks2} using path integration over the metric $h_{ij}$. Since the states do not change under spatial diffeomorphisms, one should cure the infinity arising from the volume of the gauge group as usual. Taking this point into account, the inner product can be written as 
\be\label{ip} 
\left<\Psi|\Omega\right>=\int D h_{ij} \, \Psi^*[X^\m,h_{ij}]\,\Omega[X^\m,h_{ij}]  \times \textrm{GF} \times \textrm{FP}
\ee 
where GF and FP stands for the gauge fixing and Faddeev-Popov terms (as in the standard Yang-Mills theory, this procedure should make \eq{ip} finite by removing the infinities arising from the gauge symmetry (iv)). One can show that \eq{ip} does not change under the deformation $\d X^\m=\e(y^i)n^\m$ hence the inner product is invariant under time translations, see \cite{aks2} for details. The real physical degrees of freedom can be identified as the metric $h_{ij}$ modulo coordinate transformations and obviously $|\Psi[X^\m,h_{ij}]|^2$ provides a meaningful probability distribution. 

One might naively think that the counting of degrees of freedom does not match with a spin-2 field because there are 6 independent components of $h_{ij}$ and only 3 gauge parameters. Nevertheless, the physical states must also obey $\Phi\Psi=0$ and this condition can be used to eliminate one degree of freedom. To see this, one can decompose $h_{ij}$ as 
\be\label{accp}
h_{ij}=a^2 \cc_{ij},
\ee
where $\det(\cc_{ij})=1$ (this decomposition is irreducible under spatial diffeomorphisms, i.e. $a$ and $\cc_{ij}$ do not mix with each other). The condition $\Phi\Psi=0$ gives a functional differential equation of the form 
\be
\fr{\d^2}{\d a(y^i)^2}\Psi=...
\ee
which can be used to fix the scale factor $a$ dependence of $\Psi$. In general, there are  two independent solutions (for each $a(y^i)$ variable) since  $\Phi\Psi=0$ is second order in functional derivatives but only one solution  is expected to be normalizable under \eq{ip} as $a(y^i)\to\infty$. As discussed in \cite{aks2}, the nomalizability may also require extra conditions in the theory, e.g. there must exist a negative cosmological constant in the simplest minisuperspace model. Also, it is crucial that the scale factor  ranges  in the half-line  $0<a<\infty$ otherwise for a full range $(-\infty,+\infty)$ it would be impossible to normalize the wavefunction (one should also be careful about the regularity as $a\to0$ but it looks like the path integral measure in \eq{ip} yields  enough positive powers of $a$ so that no problem arises in  that limit \cite{aks2}). 

It is instructive to see how the time evolution equation "disappears" after gauge fixing. Choosing a suitable time coordinate in the spacetime $X^\m=(t,x^i)$, one can set $t=\tau$ and $x^i=y^i$ to adopt to the conventional foliation. Next, it is possible to see that the second and third constraints in \eq{rgrc} give
\bea
&&\Psi[t,x^i+k^i,h_{ij}]=-\int d^3 x\,\left({\cal L}_{k^l} h_{ij}\right)\fr{\d\Psi}{\d h_{ij}(x)},\nn\\
&&\Psi[t+k^0,x^i,h_{ij}]=-k^0\,\int d^3 x\,\left({\cal L}_{N^l} h_{ij}\right)\fr{\d\Psi}{\d h_{ij}(x)},\label{time-tr}
\eea 
where $k^\m=(k^0,k^i)$ is a constant infinitesimal parameter and $N^i$ is the shift vector of the ADM decomposition with respect to the  coordinates $X^\m=(t,x^i)$. The right hand sides of \eq{time-tr} vanish by the fourth constraint in \eq{rgrc}, which then implies  
\be\label{trivial}
\fr{\del\Psi}{\del t}=0,\hs{10} \fr{\del\Psi}{\del x^i}=0.
\ee
Therefore, $\Psi=\Psi[h_{ij}]$ and it obeys the usual WdW equation \eq{wdw} as a consequence of the fifth constraint in \eq{rgrc}. 

In fact, \eq{trivial} is not surprising at all. Obviously any explicit dependence of $\Psi$ on the spacetime coordinates would manifestly break {\it the general covariance} and this is avoided by \eq{trivial}.  Note that this is in sharp contrast to the special relativistic covariance, where the combination of {\it coordinates} $\eta_{\m\n}x^\m x^\n$ is Lorentz invariant. No Schr\"{o}dinger equation can arise in a generally covariant theory since no such invariant combination of coordinates is possible. 

We note that the structure of the theory completely changes when one expands around a classical background solution. In \cite{ak3}, we have systematically studied how such an expansion in the action can be carried out for the perturbations around the background and showed that there appears an honest/non-vanishing Hamiltonian and additional first class constraints.  Structure-wise, the theory becomes similar to a typical Yang-Mills gauge theory (that have  infinitely many terms) and the problem of time is bypassed because the background solution provides a reference for time evolution.  In the context of the WdW equation, one can imagine a wavefunction of the form $\Psi=\Psi_{cl}\,\psi$, where $\Psi_{cl}$ is localized around the given classical background. Then, it is possible to show under certain approximations that the time independence of the full wavefunction $\del\Psi/\del t=0$ together with the WdW equation yield a Schr\"{o}dinger equation  for $\psi$ (it is also possible to utilize a WKB approximation after which a time evolution emerges with respect to the embedding coordinates viewed as a Gaussian reference fluid, see \cite{mm,mma}). As a result, there emerges a significant difference between the background independent  vs. dependent approaches to quantum gravity in terms of how time is dealt with. 

The quantization procedure proposed in this paper is by no means unique and there are obviously different approaches which may not yield equivalent results due to the nonlinear structure of general relativity. As pointed out above, {\it in certain cases} quantizing the true physical degrees of freedom after gauge fixing (deparametrization) is another viable alternative, which yields a time variable and an evolving wave-function, see e.g. \cite{ak1,ek1} for examples. It is not difficult to see that the two quantizations,  deparametrization and WdW, are {\it inequivalent}. For example, in a simple scalar field  minisuperspace model which has two variables $a$ and $\phi$, the deparametrized theory yields a wavefunction $\psi(a,\phi)$, where $a$ plays the role time (hence it is non-dynamical) and $\psi(a,\phi)$ gives a probability distribution for $\phi$ \cite{ak1}. On the other hand, in the WdW quantization as we discuss  here, the wave-function $\Psi(a,\phi)$ yields a probability distribution over the dynamical variables $a$ and $\phi$, and there is no time evolution.  In the pure minisuperspace model without any matter, the deparametrized theory is pure gauge \cite{woodard} but the WdW quantization yields a single state \cite{aks2}, again showing the inequivalence of the two methods.  It would not be surprising to see that in other approaches like loop quantum gravity \cite{ek2} or refined algebraic quantization \cite{ek3}, one  reaches different conclusions. 

Finally, in the absence of dynamical time evolution, it is not meaningful to define some of the observables such as scattering amplitudes, however one can still apply the conventional rules of quantum mechanics, for example, in calculating the expectation values of field operators in a specified state. The situation is similar to a totally isolated system which is prepared in an energy eigenstate where the Schr\"{o}dinger equation becomes trivial. Maybe not surprisingly, the canonical quantization implies that the universe should be considered as an isolated system having zero total energy. It would be interesting to develop the formalism in the presence of boundary terms, particularly for asymptotically flat spaces with finite ADM masses. In such cases, one would expect to obtain a nonzero Hamiltonian for the asymptotic fields, leading to a nontrivial time evolution of the states involving asymptotic variables. We hope to address this problem in a future publication.

\end{document}